\documentclass[conference]{IEEEtran}
\IEEEoverridecommandlockouts
\usepackage{cite}
\usepackage{subcaption}
\usepackage[hidelinks]{hyperref}
\usepackage{amsmath,amssymb,amsfonts}
\usepackage{algorithmic}
\usepackage{graphicx}
\usepackage{textcomp}
\usepackage{xcolor}
\usepackage{mathtools}
\usepackage{booktabs}
\usepackage{siunitx}
\usepackage{multirow}
\usepackage[none]{hyphenat}
\usepackage[export]{adjustbox}
\usepackage{balance}

\begin{document}

\title{Real-Time Hardware-In-the-Loop Emulation Framework for DSRC-based Connected Vehicle Applications}

\author{Ghayoor Shah$^*$, Rodolfo Valiente$^*$, Nitish Gupta$^*$, S M Osman Gani$^*$,\\
 Behrad Toghi$^*$, Yaser P. Fallah$^*$, Somak Datta Gupta$^\dagger$\\
$^*$Center for Research in Electric Autonomous Transport (CREAT), Orlando, FL\\
$^\dagger$Ford Motor Company, Dearborn, MI\\ 
 \{gshah8, rvalienter90, nitish.gupta, smosman.gani, toghi\}@knights.ucf.edu, yaser.fallah@ucf.edu, sdattagu@ford.com
        }
% \thanks{This material is based on work supported by the Ford Motor Company. Submitted to the \textit{IEEE Vehicular Technology Conference (IEEE VTC-Fall 2019)} in February 2019.}}

\maketitle

\begin{abstract}
The rapid growth of connected and automated vehicle (CAV) solutions have made a significant impact on the safety of intelligent transportation systems. However, similar to any other emerging technology, thorough testing and evaluation studies are of paramount importance for the effectiveness of these solutions. Due to the safety-critical nature of this problem, large-scale real-world field tests do not seem to be a feasible and practical option. Thus, employing simulation and emulation approaches are preferred in the development phase of the safety-related applications in CAVs. Such methodologies not only mitigate the high cost of deploying large number of real vehicles, but also enable researchers to exhaustively perform repeatable tests in various scenarios. Software simulation of very large-scale vehicular scenarios is mostly a time consuming task and as a matter of fact, any simulation environment would include abstractions in order to model the real-world system. In contrast to the simulation-based solutions, network emulators are able to produce more realistic test environments. In this work, we propose a high-fidelity hardware-in-the-loop network emulator framework in order to create testing environments for vehicle-to-vehicle (V2V) communication. The proposed architecture is able to run in real-time fashion in contrast to other existing systems, which can potentially boost the development and validation of V2V systems.

\end{abstract}
\section{Introduction}
Connected and automated vehicles (CAVs) are playing a vital role in boosting safety and efficiency in the intelligent transportation eco-system. Specifically, the connected vehicle technology has provided a medium for the development of potential cooperative safety-enhancing solutions. Connected vehicle applications employ communication technologies such as Dedicated Short-Range Communications (DSRC) \cite{jkenney:dsrcmain} and Cellular Vehicle-to-everything (C-V2X) \cite{btoghi:vnc, btoghi:vtc2019} to disseminate information for situational awareness in a cooperative manner\cite{hnmahjoub:syscon}. This shared information can be utilized for Cooperative Vehicle Safety (CVS) applications and enabling cloud-based services among the vehicles in a vehicular ad-hoc network (VANET). Given the safety-critical nature of such systems, it is essential to conduct rigorous research and extensive testing of their performance in various scenarios before vastly deploying in commercial vehicles.

 Notwithstanding the paramount importance of testing capabilities and robustness of connected vehicle systems, it is yet a challenging problem to materialize a controlled environment for the purpose of performing repeatable tests in the desired vehicular scenarios. Performing repeatable real-world field tests for all possible mobility and communication situations in different scenarios is neither technically feasible nor cost-efficient. This fact is mainly due to the highly-dynamic and stochastic nature of vehicular communications and in general any communication system, alongside a multitude of other contributing factors. As an instance, in a large-scale vehicular test with potentially hundreds of vehicles, tracking a specific incident which may happen very rarely may be challenging and even if it becomes possible, repeating that incident in the exact same environmental condition in order to further study the phenomena can be burdensome.

Given the aforementioned complications of field trials for vehicular scenarios especially in high-risk and near-crash cases, there is a tendency to employ solutions based on simulations as an alternative to impractical and costly field tests. However, any simulation effort contains some level of abstraction in order to model real phenomena, e.g., radio propagation channels, electronic transceiver devices, traffic flow, etc. Thus, fidelity and reliability of the simulation tools always depend on the accuracy of the abstracted models. In addition, simulating realistic large-scale scenarios by utilizing current primary simulation frameworks such as NS-3, OMNET++, and OPNET is a very time-consuming task. As an example, we particularly show later in this paper that NS-3 can take orders of magnitude longer than the real simulation period. 

In an attempt to address the former, authors in \cite{gupta2012bsm} have proposed a network emulator for VANET applications. This emulator creates a virtual cloud of Basic Safety Messages (BSMs) \cite{sae:j2735} as if they are broadcasted by hundreds of vehicles. An early study in \cite{bansal2011cross} has also cross referenced the results from a DSRC emulation platform and NS-2 simulations to prove the accuracy of its simulation environment. Another existing work in the literature is the high-fidelity DSRC simulator proposed in \cite{ypfallah:bookchapter} which is claimed by authors to be able to efficiently produce realistic results for VANET simulations. However, all of the above-mentioned solutions experience very long run-times and hence are not able to operate in a real-time fashion.

In this work, we propose a real-time high-fidelity hardware-in-the-loop remote vehicle emulation framework (RVE). Our proposed architecture can be employed to clone the behavior of potentially thousands of connected vehicles and create artificial channel congestion required for the purpose of performance and scalability studies in VANETs. As we show in Section IV, our proposed solution is not only able to produce results as accurately as those by NS-3, but it is also faster by around two orders of magnitude. The main contributions of this work are the real-time capability and a more realistic implementation since we use actual DSRC modems as a part of our architecture. The aforementioned features make our proposed architecture a desirable alternative for large vehicle deployments in field trials. We have integrated models of important contributing phenomena such as the hidden node problem to ensure the accuracy and reliability of our results. Such a cost-efficient platform enables researchers to conduct repeatable and controlled tests for CAV applications.  In this work, we focus on DSRC as the underlying communication technology, however, our proposed architecture can also be modified to work on top of C-V2X or other communication technologies. 

The rest of the paper is organized as follows. Section II provides an overview of Vehicle-to-Vehicle (V2V) communication and DSRC technology. Section III discusses our proposed architecture and implementation details alongside a description of the two main components of the architecture, namely the mobility log generator and Real-Time Communication Simulator (RTCSim). In section IV, the performance of our proposed framework is evaluated and compared to the results from the NS-3 simulator.  Finally, we conclude the paper in Section V by providing a summary and potential research directions.

\section{Background}
Multiple comprehensive studies and survey papers exist in the literature which investigate V2V communication and DSRC technology in particular \cite{jkenney:dsrcmain, hnmahjoub:cavs, hnmahjoub:vtc}. Although the details of V2V communication and DSRC are out of the scope of this paper, for the sake of completeness, we provide a brief overview of both of the topics in this section.

In a CVS system, Host Vehicle ($HV$) and Remote Vehicles ($RVs$) utilize an ad-hoc wireless network in order to share their situational awareness with each other. Connected vehicles are equipped with a V2V on-board unit (OBU) which enables them to establish V2V communication or Vehicle-to-infrastructure (V2I) communication through road-side units (RSUs). Each DSRC OBU mainly consists of a general-purpose processor, IEEE802.11p \cite{5514475} radio front-end, a positioning module, i.e., Global Navigation Satellite System (GNSS), in addition to a handful of interfaces for obtaining vehicle data, e.g., Controller Area Network (CAN) bus. Safety applications use the collected data from OBU interfaces in the form of BSMs \cite{sae:j2735} in order to predict a potential hazard or a near-crash scenario and subsequently alert the driver \cite{liu2016delivering}.

In the IEEE 802.11 protocol, the fundamental mechanism to access the medium is called the Distributed Coordination Function (DCF). This is a random access scheme based on the Carrier Sense Multiple Access with Collision Avoidance (CSMA/CA) protocol. The work in \cite{bianchi1996performance} provides more information on the IEEE 802.11 protocol. Within IEEE 802.11, DSRC is known as IEEE 802.11p, which is an amendment to the overall IEEE 802.11 standard and essentially is IEEE 802.11a adjusted to work in 10MHz channels at 5.9GHz frequency band. DSRC is a two-way short-to-medium-range wireless communication technology that permits a high data transmission which is critical in communication-based active safety applications \cite{jkenney:dsrcmain}. In recent years, there has been a considerable interest in DSRC research and several research efforts have been made in DSRC based communication \cite{gani2016high}. This promising development is designed to support V2V and V2I communication that would enable advanced active vehicle safety among other safety applications \cite{ieee2016ieee}. 

\section{Proposed Solution}
One of the paramount benefits of the proposed RVE framework is that it supports the incorporation of the desired vehicle mobility pattern or traffic scenario, thereby broadening the testing possibility to scenarios that might be too dangerous or costly to be conducted or reproduced in a real-world field trial. As an instance, near-crash or large-scale vehicular scenarios are very burdensome to create or repeat in a real-world field test setup. However, with RVE, vehicle mobility traces for such extensive and complex scenarios can be generated using either real-world field tests or simulation results or a combination of both possibilities. Mobility traces can then be fed to RTCSim to reconstruct the scenario for the $HV$. The standardized over-the-air messages in the scenario representing the emulated $RVs$ are received by the $HV$ as if they have been broadcasted by real $RVs$ in its proximity.
\subsection{System-level Architecture}
From a system level perspective, RVE consists of two main sub-systems, namely the mobility log generator and RTCSim. While the former is responsible for generating vehicle movement trajectories that are logged and fed to the RTCSim module, the latter simulates the behavior of MAC and PHY layers and generates BSMs to create an environment of emulated $RVs$. We present a more detailed description of these modules and our proposed design in this section.
\\
\begin{figure}[t]
\centerline{\includegraphics[width=.45\textwidth]{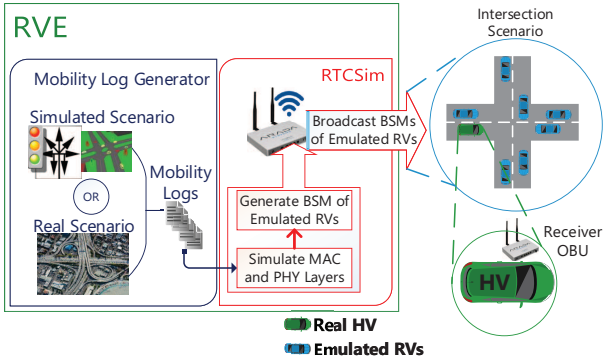}}
\caption{RVE Framework}
\label{fig1}
\end{figure}

\subsubsection{Mobility Log Generator}
Figure \ref{fig1} illustrates a system-level overview of the framework. Emulated $RVs$ are required to follow a pre-configured trajectory while the safety applications are being tested. Thus, we require a method to generate these trajectory logs which can then be fed to RTCSim. There can be two different approaches for creating these trajectory logs. The fist approach involves employing real vehicles in a controlled environment to record their travelled path and the second utilizes a simulation-based mobility handler framework. Although mobility logs recorded using the first method are highly accurate, producing them is a time-consuming task due to the complexities of on-field experiments. Moreover, testing all possible corner cases of safety applications requires repeatability of experiments in which case generating mobility logs from real vehicles turns out to be inefficient and costly. 

In order to generate trajectory logs using the second method, we use a traffic simulator called SUMO. The road scenario to be tested can be exported from a map API such as OpenStreetMap and fed to the SUMO simulator. SUMO then generates a mobility log readable by NS-3 network simulator for a given vehicle density and road scenario. The log contains map-sharing information such as time-stamps, vehicle IDs, GPS positions, velocities and other required data fields. The mobility log is split into a separate log file for each vehicle and then these multiple log files are used by RTCSim to emulate communication between them. Based on the transmission frequency of the vehicles, each vehicle log file contains a certain number of BSM packets for different time-stamps.
\\
\subsubsection{RTCSim}
RTCSim is the second and the core module of RVE architecture. It creates a virtual cloud of emulated $RVs$ in the neighborhood of the $HV$ in which vehicles are connected together via the V2V link. The earlier discussed mobility log generator module feeds RTCSim with the required mobility traces enabling it to emulate the real-time MAC and PHY layer behavior of the emulated $RVs$ from the perspective of the $HV$. We delve deeper into the implementation details of RTCSim in the remainder of this section.
\subsection{Implementation Details}
\begin{figure}[t]
\centerline{\includegraphics[width=.45\textwidth]{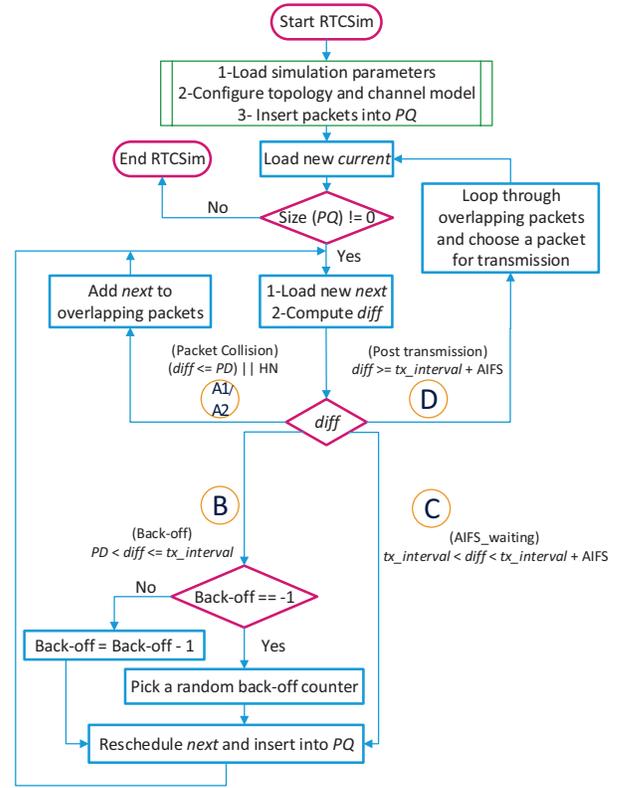}}
\caption{RTCSim Flowchart}
\label{fig2}
\end{figure}

Figure \ref{fig2} demonstrates the flow-chart representation of the logic that is implemented in the RTCSim module. At core, an RTCSim data packet sent by an emulated RV encapsulates BSM data. In order to begin the emulation of $N$ number of vehicles, the simulation parameters presented in table \ref{table:configs} are loaded and a priority queue ($PQ$) of size $N$ is initialized with one packet from each vehicle. The $PQ$ is sorted based on the scheduled packet generation time-stamps in a way that the first scheduled packet is always at the top of the queue. As the simulation proceeds, the packet at the top of the queue ($current$) is popped and considered for transmission. RTCSim then iteratively goes through each of the overlapping packets ($next$) and calculates the time difference ($diff$) between $current$ and each $next$ packet. Based on $diff$, the algorithm goes into one of the five possible states, namely $A1$, $A2$, $B$, $C$, and $D$ as mentioned in the flowchart in figure \ref{fig2}. Each state represents a mode that can be experienced in the CSMA/CA scheduling procedure. All the states are mentioned in detail below.
\\
\subsubsection{A1. Collision\_PD (Packet collision due to Propagation Delay)}
After a vehicle starts transmitting a packet on the channel, it takes a short interval of time for other vehicles in the network to be able to sense the channel as busy. This interval of time is known as the Propagation Delay ($PD$). RTCSim arrives in this condition when $diff<=PD$. If $current$ packet starts transmitting on the channel and $next$ has a packet generation time-stamp within the range of the $PD$ of $current$, $next$ would not be able to sense the transmission of $current$ and would consider the channel as idle. As a result, there would be a packet collision between $current$ and $next$ and $next$ would then be added to the list of overlapping packets to $current$. 
\\
\subsubsection{A2. Collision\_HN (Packet collision due to Hidden Node)}
RTCSim enters this condition whenever $next$ packet is a Hidden Node (HN) to $current$ packet. As opposed to A1, the A2 condition can occur throughout the transmission of $current$ and not just within its $PD$. If the packet generation time-stamp of $next$ is within the transmission of $current$, it would lead to a packet collision since $next$, being hidden to $current$, would sense the channel to be idle and would begin its transmission. As a result, $next$ would be added to the list of overlapping packets.
\\
\subsubsection{B. Back-off}
The back-off condition occurs when $PD<diff<=tx\_interval$, where $tx\_interval$ is the duration of the $current$ packet on the channel. Since the $diff$ value is greater than $PD$ of $current$ and $next$ is not hidden, this means that $next$ can sense the transmission of $current$ in the channel. Thus, $next$ would go through the back-off procedure shown in the flowchart in figure \ref{fig2} in order to reschedule its packet generation time-stamp and then get re-inserted and re-sorted in the $PQ$. When the back-off counter of that packet reaches zero, the packet is scheduled to transmit at the next available idle slot after the current transmission.
\\
\subsubsection{C. AIFS\_waiting}
Arbitration Inter-Frame Space (AIFS) is the idle interval of time on the channel after the transmission of every packet. The AIFS\_waiting condition shown in figure \ref{fig2} occurs if $tx\_interval < diff <= tx\_interval + AIFS$. The design rationale behind AIFS interval is to ensure that the channel is kept idle for a short interval of time after each transmission. Equation \ref{equ:AIFSdefinition} formulates the definition of AIFS,
\begin{equation} \label{equ:AIFSdefinition}
\textit{AIFS} = \textit{SIFS} + 2\times \textit{slot\_time}
\end{equation}
where SIFS is Short Inter-Frame Space.
AIFS interval prevents the backed-off packets from starving to transmit in the case where all packets are rescheduled one after the other without any idle time among them. Hence, after each packet transmission, at least one idle slot would be observed by all the packets. If $next$ arrives during the AIFS interval right after the transmission of $current$ packet, $next$ would be rescheduled to arrive after the AIFS interval, thereby keeping the AIFS interval idle.
\\
\subsubsection{D. Post-Transmission}
This state happens in the case when $diff>tx\_interval+AIFS$. Since $diff$ between $current$ and $next$ is greater than the $tx\_interval$ of $current$ and the AIFS interval following it, this implies that $next$ is not interfering with $current$ in this condition. Thus, we do not need to reschedule $next$ to ensure successful transmission of $current$. Therefore, in this condition we just need to focus on the overlapping packets to $current$ stored in previous iterations. We iterate through all the overlapping packets and after modelling the capture effect, we choose one of them for transmission.
\section{Analysis and Results}
%
% Intel® Core™ i7-6700 CPU @ 3.40GHz × 8 
It is important to validate RTCSim against a well-established baseline in order to test its accuracy. For this purpose, we leverage NS-3 simulator which is a widely used network simulation tool for CAV applications. We use a Linux personal computer with Core i7-6700 processor and 32GB of RAM to run the simulations for both NS-3 and RTCSim. The results of NS-3 for V2V communication have already been validated for accuracy in \cite{ypfallah:bookchapter}. Although NS-3 is unsuitable for real-time emulation-based testing due to its high communication latency and non-real-time performance, it is still a high-fidelity simulation tool for CAV systems. Thus, a comparison between NS-3 and RTCSim simulation results could provide a sufficient validation for the behavior and accuracy of RTCSim. It should be noted that it is essential to configure NS-3 and RTCSim with the same parameters to ensure a meaningful comparison. The parameter settings used for both NS-3 and RTCSim are shown in Table \ref{table:configs}.
\begin{table}[htbp]
\centering
\caption{Parameters \& Configurations for NS-3 and RTCSim}
\begin{center}
\bgroup
\def\arraystretch{1.4}
\begin{tabular*}{0.32\textwidth}{@{\extracolsep{\fill} }  l r }
\hline
\hline
Contention Window Size                  & $[0-15]$\\
Packet Transmission Rate                 & $10\text{Hz}$\\
Transmission Power                         & $20 \text{dBm}$\\
Slot Time                           & $13 \mu s$\\
Simulation Time                       & $20s$\\
%\# of Packets per Vehicle               & $200$\\
\hline
\end{tabular*}
\egroup
\label{table:configs}
\end{center}
\end{table}

For the purpose of evaluating the performance of our proposed solution, we utilize two key performance metrics, namely Channel Busy Percent (CBP) and Packet Error Rate (PER). We consider two different channel models i.e. Three Log Distance Propagation Loss Model (referred to as \textbf{T}) and Fowlerville Propagation Loss Model (referred to as \textbf{F}). For each channel model, we use three different topologies i.e. a disk representing a roundabout with radius of 500m  (referred to as \textbf{D}), a 3000m  long linear road (referred to as \textbf{L}) and an intersection with two perpendicular bisecting roads of 1500m length (referred to as \textbf{I}). Therefore in total, we consider six different scenarios (\textbf{FD}, \textbf{FL}, \textbf{FI}, \textbf{TD}, \textbf{TL}, and \textbf{TI}) to compare PER and CBP of NS-3 and RTCSim. 

\begin{figure}[htbp]
\centerline{\includegraphics[width=.48\textwidth]{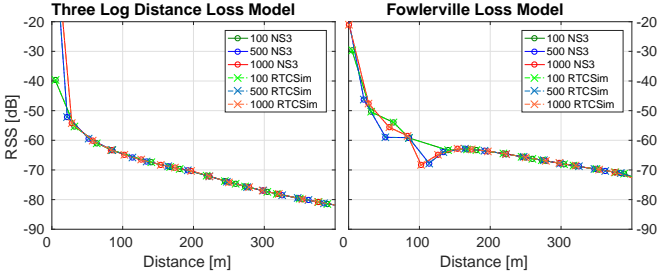}}
\caption{RSS comparison of NS-3 and RTCSim.}
\label{fig3}
\end{figure}

As an initial sanity check to make sure that both channel models perform exactly the same way in NS-3 and RTCSim, we compare the Received Signal Strength (RSS) versus distance between $HV$ and emulated $RVs$ for both models as shown in figure \ref{fig3}. It can be observed that the RSS plots of both the channel models in NS-3 and RTCSim completely match with each other. For each of the six defined test scenarios, we conduct our study in low (100 vehicles), medium (500 vehicles) and high (1000 vehicles) densities. For each topology, the $HV$ and emulated $RVs$ are randomly placed following a uniform distribution and are moving with constant speed.

\begin{figure}[htbp]
\centerline{\includegraphics[width=.48\textwidth]{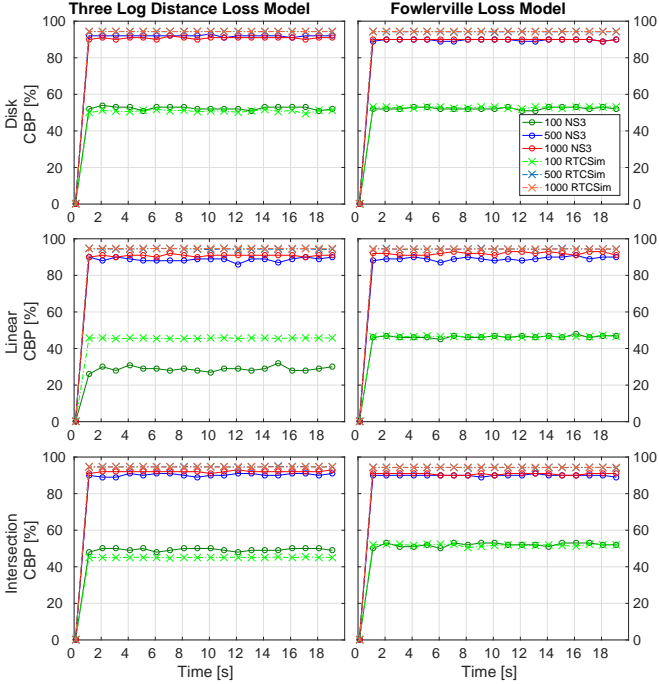}}
\caption{CBP comparison of NS-3 and RTCSim}
\label{fig4}
\end{figure}

\begin{figure}[htbp]
\centerline{\includegraphics[width=.48\textwidth]{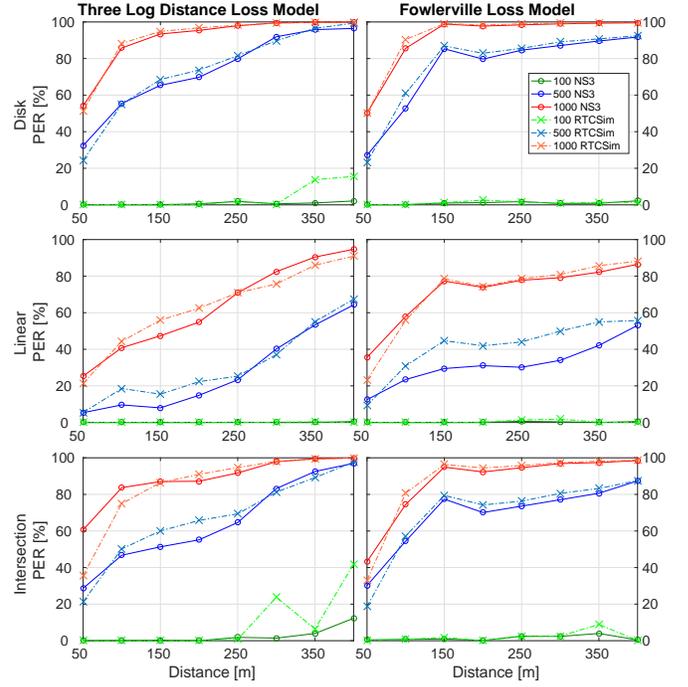}}
\caption{PER comparison of NS-3 and RTCSim}
\label{fig5}
\end{figure}
Figures \ref{fig4} and \ref{fig5} illustrate the CBP and PER plots for all the six scenarios mentioned earlier where each plot provides a comparison between NS-3 and RTCSim for low, medium and high vehicle densities. For every scenario, CBP is plotted against simulation time (20s) and PER is plotted against distance between $HV$ and emulated $RVs$. The distance axis in the PER plots is limited to 400m since we do not need to focus on distances larger than that for CVS applications.

It can be noticed that the CBP and PER plots of all scenarios show a promising similarity between NS-3 and RTCSim results. For every plot in figure \ref{fig4}, it can be seen that an increase in vehicle density causes the channel to become busier and thus possess a higher CBP. This is the reason that the CBP for high-vehicle density is higher than that of medium-vehicle density which, in turn is higher than that of low-vehicle density for all plots. For each vehicle density in every plot in figure \ref{fig5}, an increase in distance causes the RSS to reduce, which creates more chances of hidden nodes in the network and consequently a higher probability of packet collisions, thus a higher PER. This effect is clearly visible for medium and high-vehicle densities because they already have a high CBP as can be viewed in figure \ref{fig4} and as the distance increases, it causes more packet collisions and thus, an increasing PER. On the contrary, this effect is minimal in the case of low vehicle densities since they have low CBP and as the distance increases, it doesn't result in a noticeable increase in the number of collisions. Therefore, the PER for low-vehicle densities in all the plots remains low.

Tables \ref{table:CBPcomparison} and \ref{table:PERcomparison} provide tabular versions of CBP and PER of NS-3 and RTCSim by averaging the CBP and PER plots for all the scenarios in figures \ref{fig4} and \ref{fig5}. It can be noted that the average CBP and PER values of RTCSim are very similar to those of NS-3 for all scenarios. The slight differences are due to the abstractions in RTCSim as well as the inevitable stochasticity of a wireless communication channel. 

\begin{table}[t]
\centering
\caption{COMPARISON OF AVERAGE CBP OF NS-3 AND RTCSim}
\begin{center}
\begin{tabular}{|c||c||c|c|c|c|}
\hline
\multicolumn{6}{|c|}{Average CBP (\%)}                                                                \\ \hline
\multirow{2}{*}{Topology} & \multirow{2}{*}{Vehicles} & \multicolumn{2}{c|}{\textit{Fowlerville}} & \multicolumn{2}{c|}{\textit{Three Log-distance}} \\ \cline{3-6} 
                                &       & \textbf{RTCSim}   & NS-3      & \textbf{RTCSim}       & NS-3      \\ \hline \hline
\multirow{3}{*}{Disk}           & 100   & \textbf{ 52.62}   & 52.40      & \textbf{ 51.03}       & 52.27     \\ \cline{2-6} 
                                & 500   & \textbf{ 94.21}   & 89.76     & \textbf{ 94.28}       & 91.94     \\ \cline{2-6} 
                                & 1000  & \textbf{ 94.36}   & 89.96     & \textbf{ 94.37}       & 90.69     \\ \hline \hline
\multirow{3}{*}{Linear}         & 100   & \textbf{ 46.79}   & 46.68     & \textbf{ 45.60}       & 29.30     \\ \cline{2-6} 
                                & 500   & \textbf{ 94.39}   & 89.03     & \textbf{ 94.38}       & 89.02     \\ \cline{2-6} 
                                & 1000  & \textbf{ 94.43}   & 92.18     & \textbf{ 94.67}       & 90.69     \\ \hline \hline
\multirow{3}{*}{Intersection}   & 100   & \textbf{ 51.72}   & 52.15     & \textbf{ 45.09}       & 49.40     \\ \cline{2-6} 
                                & 500   & \textbf{ 94.24}   & 90.01     & \textbf{ 94.65}       & 90.50     \\ \cline{2-6} 
                                & 1000  & \textbf{ 94.36}   & 90.55     & \textbf{ 94.58}       & 91.89     \\ \hline \hline
\end{tabular}
\end{center}
\label{table:CBPcomparison}
\end{table}
\begin{table}[t]
\centering
\caption{COMPARISON OF AVERAGE PER OF NS-3 AND RTCSim}
\begin{center}
\begin{tabular}{|c||c||c|c|c|c|}
\hline
\multicolumn{6}{|c|}{Average PER (\%)}                                                                \\ \hline
\multirow{2}{*}{Topology} & \multirow{2}{*}{Vehicles} & \multicolumn{2}{c|}{\textit{Fowlerville}} & \multicolumn{2}{c|}{\textit{Three Log-distance}} \\ \cline{3-6} 
                                &       & \textbf{RTCSim}   & NS-3      & \textbf{RTCSim}       & NS-3      \\ \hline \hline
\multirow{3}{*}{Disk}           & 100   & \textbf{ 1.08 }   & 0.82      & \textbf{ 0.52 }       & 0.55      \\ \cline{2-6} 
                                & 500   & \textbf{ 71.51}   & 69.44     & \textbf{ 65.52}       & 65.76     \\ \cline{2-6} 
                                & 1000  & \textbf{ 89.28}   & 88.28     & \textbf{ 88.13}       & 87.68     \\ \hline \hline
\multirow{3}{*}{Linear}         & 100   & \textbf{ 0.56}   & 0.16    & \textbf{ 0.03}       & 0.02     \\ \cline{2-6} 
                                & 500   & \textbf{ 36.80}   & 26.75     & \textbf{ 20.6}       & 16.87    \\ \cline{2-6} 
                                & 1000  & \textbf{ 65.26}   & 66.93    & \textbf{ 55.1}       & 53.2    \\ \hline \hline
\multirow{3}{*}{Intersection}   & 100   & \textbf{ 1.16 }   & 1.21      & \textbf{ 4.21 }       & 0.59      \\ \cline{2-6} 
                                & 500   & \textbf{ 64.00}   & 63.79     & \textbf{ 57.86}       & 54.94     \\ \cline{2-6} 
                                & 1000  & \textbf{ 82.94}   & 82.74     & \textbf{ 80.21}       & 84.69     \\ \hline \hline
\end{tabular}
\end{center}
\label{table:PERcomparison}
\end{table}

One main premise over which RTCSim proves to be a powerful tool for emulation is its low-latency performance. Table \ref{table:ExecutionTime} shows a comparison of mean execution time of all above experiments in NS-3 and RTCSim. It can be observed that execution time of NS-3 for all vehicle densities is higher than the simulation time. This proves the argument made earlier in the paper that NS-3 is not useful for real-time CVS applications. On the contrary, RTCSim is not just faster than NS-3 but it is also faster than real-time for all the scenarios studied in this paper. Thus, RTCSim not only models the communication behavior of the desired number of vehicles accurately and has minimum latency as compared to NS-3 but it also allows real-time emulation.  

\begin{table}[t]
\centering
\caption{COMPARISON OF MEAN EXECUTION TIME OF NS-3 AND RTCSim}
\begin{center}
\begin{tabular}{|c||c|c|}
\hline
\multicolumn{3}{|c|}{Mean execution time (s)}                                                                \\ \hline
%\multirow{2}{*}{\# of Vehicles} & \multirow{2}{*}{NS-3} & %\multirow{2}{*}{\textbf{RTCSim} } 
Number of Vehicles & NS-3  & \textbf{RTCSim}  \\ [0.5ex]  \hline
 \hline
100  & 266.41                   &    \textbf{0.27}    \\ \hline 
500 & 792.63                  &     \textbf{2.43}  \\ \hline 
1000& 2175.60                   &      \textbf{6.79}\\ \hline
\hline 

\end{tabular}
\end{center}
\label{table:ExecutionTime}
\end{table}
%
%This paper presents an easily configurable real-time RVE framework to test DAS %safety applications. A main component of the RVE is the communication model that %allows accurately emulating the behavior of up to thousand vehicles in real-time. %This allows hardware in the loop testing for CAV applications in scenarios with a %large number of vehicles, without needing to setup large field tests. The %proposed RVE architecture is robust to various channel models, topologies and %number of vehicles. Inclusion of models for hidden nodes and capture effect serve %as major contributions in this architecture. While simulators such as NS-3 cannot %operate in real-time for networks with more than 50 vehicles, the proposed RVE %emulator has been tested to maintains real-time performance for 1000-vehicle %networks, without noticeable loss of accuracy. 

%The framework can be further extended to improve the channel performance by using %the distance dependent congestion control strategies based on parameters like %power, rate and/or message content. The architecture can also be modified to %support testing of LTE-based safety application. 
%

\section{Concluding Remarks}
%
%As any other emerging technology, extensive testing and thorough study of connected vehicle technology is of great importance. The safety-critical nature of this technology requires the contributing developers to thoroughly investigate the high-risk near-crash cases as well as large-scale vehicular scenarios. However, conducting real-world field trials for such test cases is a complex time-consuming task and seems to be not a feasible, or at least an efficient solution. Hence, software-based solutions based on simulation and emulation mediums have been proposed in order to address the above-mentioned need. The main players in the simulation domain, namely NS-3 and OMNET++ simulators, operate in orders of magnitude longer run-times, compared to the actual simulation time. Hence, employing such simulation environments makes the development phase very time-consuming and slows down the pace of the safety-critical investigations.%

In this work, we propose an emulation framework for CAV applications based on the DSRC protocol stack which is able to operate in a real-time fashion. Through analysis in different channel models and vehicular topologies, we prove the high fidelity of our proposed solution with  a significant improvement in the run-time as compared to that for NS-3. A main component of our RVE is the communication model that allows accurate emulation of the behavior of up to 1000 vehicles in real-time. This allows hardware-in-the-loop testing for CAV applications in scenarios with a large number of vehicles without needing to set up large field tests. Consideration of the hidden node problem and the capture effect also serve as contributions in this architecture. While simulators such as NS-3 cannot operate in real-time, the proposed RVE emulator has been tested to maintain real-time performance without a noticeable loss in accuracy.

The framework can be further extended to improve the channel performance by using distance dependent congestion control strategies based on parameters like power, rate and/or message content. Another future research direction could be modifying our proposed solution to operate on other alternative vehicular communication technologies such as C-V2X and also adding more complex applications, e.g., SAE J2945/1 to the protocol stack.
\balance
\bibliography{refs.bib}{}

\begin{thebibliography}{10}

\bibitem{jkenney:dsrcmain}
J.~B. Kenney.
\newblock Dedicated short-range communications (dsrc) standards in the united
  states.
\newblock {\em Proceedings of the IEEE}, 99(7):1162--1182, July 2011.

\bibitem{btoghi:vnc}
B.~Toghi, M.~Saifuddin, H.~N. Mahjoub, M.~O. Mughal, Y.~P. Fallah, J.~Rao, and
  S.~Das.
\newblock Multiple access in cellular v2x: Performance analysis in highly
  congested vehicular networks.
\newblock In {\em 2018 IEEE Vehicular Networking Conference (VNC)}, pages 1--8,
  Dec 2018.

\bibitem{btoghi:vtc2019}
Behrad {Toghi}, Md~{Saifuddin}, Yaser~P. {Fallah}, and M.~O. {Mughal}.
\newblock {Analysis of Distributed Congestion Control in Cellular
  Vehicle-to-everything Networks}.
\newblock {\em arXiv e-prints}, page arXiv:1904.00071, Mar 2019.

\bibitem{hnmahjoub:syscon}
Hossein {Nourkhiz Mahjoub}, Behrad {Toghi}, S~M {Osman Gani}, and Yaser~P.
  {Fallah}.
\newblock {V2X System Architecture Utilizing Hybrid Gaussian Process-based
  Model Structures}.
\newblock {\em arXiv e-prints}, page arXiv:1903.01576, Mar 2019.

\bibitem{gupta2012bsm}
Somak~Datta Gupta, Chih-Che Lin, and Ching-Yao Chan.
\newblock Bsm emulator-advanced vehicle safety application testbed.
\newblock In {\em 2012 15th International IEEE Conference on Intelligent
  Transportation Systems}, pages 1868--1873. IEEE, 2012.

\bibitem{sae:j2735}
SAE International.
\newblock Dedicated short range communications (dsrc) message set dictionary.
\newblock Standard Doc J2735, Society of Automotive Engineers, 03 2016.

\bibitem{bansal2011cross}
Gaurav Bansal, John~B Kenney, and Aaron Weinfield.
\newblock Cross-validation of dsrc radio testbed and ns-2 simulation platform
  for vehicular safety communications.
\newblock In {\em 2011 IEEE Vehicular Technology Conference (VTC Fall)}, pages
  1--5. IEEE, 2011.

\bibitem{ypfallah:bookchapter}
Yaser~P. Fallah and S.~M. Osman~Gani.
\newblock {\em Efficient and High Fidelity DSRC Simulation}, pages 217--243.
\newblock Springer International Publishing, Cham, 2019.

\bibitem{hnmahjoub:cavs}
H.~N. {Mahjoub}, B.~{Toghi}, and Y.~P. {Fallah}.
\newblock A driver behavior modeling structure based on non-parametric bayesian
  stochastic hybrid architecture.
\newblock In {\em 2018 IEEE 88th Vehicular Technology Conference (VTC-Fall)},
  pages 1--5, Aug 2018.

\bibitem{hnmahjoub:vtc}
H.~N. {Mahjoub}, B.~{Toghi}, and Y.~P. {Fallah}.
\newblock A stochastic hybrid framework for driver behavior modeling based on
  hierarchical dirichlet process.
\newblock In {\em 2018 IEEE 88th Vehicular Technology Conference (VTC-Fall)},
  pages 1--5, Aug 2018.

\bibitem{5514475}
Ieee standard for information technology-- local and metropolitan area
  networks-- specific requirements-- part 11: Wireless lan medium access
  control (mac) and physical layer (phy) specifications amendment 6: Wireless
  access in vehicular environments.
\newblock {\em IEEE Std 802.11p-2010 (Amendment to IEEE Std 802.11-2007 as
  amended by IEEE Std 802.11k-2008, IEEE Std 802.11r-2008, IEEE Std
  802.11y-2008, IEEE Std 802.11n-2009, and IEEE Std 802.11w-2009)}, pages
  1--51, July 2010.

\bibitem{liu2016delivering}
Jun Liu and Asad~J Khattak.
\newblock Delivering improved alerts, warnings, and control assistance using
  basic safety messages transmitted between connected vehicles.
\newblock {\em Transportation research}, 2016.

\bibitem{bianchi1996performance}
Giuseppe Bianchi, Luigi Fratta, and Matteo Oliveri.
\newblock Performance evaluation and enhancement of the csma/ca mac protocol
  for 802.11 wireless lans.
\newblock In {\em Proceedings of PIMRC'96-7th International Symposium on
  Personal, Indoor, and Mobile Communications}, volume~2, pages 392--396. IEEE,
  1996.

\bibitem{gani2016high}
SM~Osman Gani, Amin Tahmasbi-Sarvestani, Mohammad Fanaei, and Yaser~P Fallah.
\newblock High fidelity dsrc receiver model for ns-3 simulation using
  large-scale field data.
\newblock In {\em 2016 IEEE Wireless Communications and Networking Conference},
  pages 1--6. IEEE, 2016.

\bibitem{ieee2016ieee}
IEEE 1609~Working Group et~al.
\newblock Ieee standard for wireless access in vehicular environments-security
  services for applications and management messages.
\newblock {\em IEEE Std}, pages 1609--2, 2016.

\end{thebibliography}
\bibliographystyle{unsrt}
% \end{multicols}
\end{document}